\documentclass[showpacs,prl,aps,superscriptaddress,floatfix,tightenlines,twocolumn]{revtex4}

\usepackage{amsmath}
\usepackage{amssymb}
\usepackage{amsfonts}
\usepackage{epsfig}
\usepackage{subfigure}
\usepackage[dvips]{color}
\usepackage{bm}

\begin{document}

\title{Criterion on remote clocks synchronization within a Heisenberg
scaling accuracy}
\author{Yong-Liang Zhang}
\affiliation{School of Physics, Peking University, Beijing 100871, China}
\author{Yu-Ran Zhang}
\affiliation{Institute of Physics, Chinese Academy of Sciences, Beijing 100190, China}
\author{Liang-Zhu Mu}
\email{muliangzhu@pku.edu.cn}
\affiliation{School of Physics, Peking University, Beijing 100871, China}
\author{Heng Fan}
\email{hfan@iphy.ac.cn}
\affiliation{Institute of Physics, Chinese Academy of Sciences, Beijing 100190, China}

\date{\today}
\pacs{03.67.Hk, 06.30.Ft, 95.55.Sh, 03.67.Mn}
\begin{abstract}
We propose a quantum method to judge whether two spatially separated
clocks have been synchronized within a specific accuracy $\sigma$.
If the measurement result of the experiment is obviously a nonzero
value, the time difference between two clocks is smaller
than $\sigma$; otherwise the difference is beyond $\sigma$.
On sharing the 2$N$-qubit bipartite maximally entangled state in
this scheme, the accuracy of judgement can be enhanced to
$\sigma\sim{\pi}/{(\omega(N+1))}$. This criterion is consistent
with Heisenberg scaling that can be considered as beating standard
quantum limit, moreover, the unbiased estimation
condition is not necessary.
\end{abstract}
\maketitle

\emph{Introduction.--}
Quantum entanglement, a distinctive feature of quantum mechanics,
is at the heart of applications in distributed
systems, e.g., quantum key distribution and clock synchronization.
Clock synchronization with high precision is a fundamental and
an important problem in that it is crucial for many modern
technologies and researches, such as global positioning system
(GPS), long baseline interferometry, synchronous data transfer,
gravitational wave observation (LIGO), tests of theory of
general relativity, and distributed computation. There are two standard
methods for synchronizing two spatially separated clocks in the
frame of special theory of relativity. One is based on Einstein's
synchronization procedure which uses an operational line-of-sight
exchange of light pulses between two spatially separated clocks
\cite{einstein}. The other one is based on the internal time
evolution of quantum systems, like Eddington's infinitesimally
slow clock transport \cite{eddington}. The quantum clock
synchronization method based on the strength of sharing prior entanglement
has also been proposed in \cite{jozsa}, and has been generalized to
several multiparty clock synchronization protocols
\cite{krco,ben-av,ren}. Independent of
the parties's knowledge of their relative locations or of the
properties of the intervening media,
these procedures utilize the instantaneity of wavefunction
collapsing after the measurement is performed on the shared
entangled states. Since the process of distributing entanglement
is adiabatic, these protocols are tantamount to Eddington
protocol. There are some relevant experimental works done on the
quantum clock synchronization protocols: an experiment focusing
on the quantum clock synchronization implementation has also been
reported in Ref.~\cite{APL}; in addition the progressive techniques
of multi-photon entanglement \cite{mpe} are basic and promising for
the realization of these protocols.

Quantum entanglement-enhanced parameter estimation that plays
a vital role in quantum metrology uses quantum mechanical
property to enhance the sensitivity of the measurement of classical
quantities. It has been pointed that the standard quantum limit
${1}/\sqrt{N}$, where $N$ is the number of particles used in the
measurement, can be beaten by using the coherent light with squeezed
vacuum \cite{Phys. Rev. D 23.1693}. In the study of quantum metrology,
quantum Fisher information theory and quantum Cram\'{e}r-Rao bound
based on the statistical distance of states have been proposed
and developed in
\cite{fisher,cramer,holevo,Braunstein,caves,Escher}. The NOON
state has been demonstrated to be able to achieve a phase
sensitivity saturating the Heisenberg limit ${1}/{N}$
\cite{bollinger}. Some related strategies have been proposed to
perform a high precision in quantum metrology framework
\cite{Giovannetti3,Giovannetti1,Giovannetti2}, whilst many
experiments have also been performed on this topic
\cite{Nagata,higgins,Kacprowicz,afek,xiang,Yonezawa}.
One direct and natural idea to apply this technique is in the
clock synchronization. Chuang \cite{chuang} has presented a high
efficiency quantum ticking qubits handshake protocol which allows
two remote clocks to be synchronized independent of message
transport time; and a similar protocol has been proposed to beat
the standard quantum limit \cite{burgh}.

In this Letter, we relate the quantum clock synchronization
protocol to the problem of estimating an unknown parameter.
We investigate the performance of the bipartite maximally entangled
spin-zero singlet in the scheme of two-clock synchronization
and offer a standard to judge whether two spatially
separated clocks have been synchronized in a specific accuracy.
This criterion is practical and is consistent to Heisenberg scaling,
additionally it does not rely on the unbiased estimation condition which is a
fundamental hypotheses in the quantum Fisher information theory.

\emph{General framework of quantum clock synchronization.--}
Suppose two spatially separated parties, Alice and Bob, rested on
the same reference frame, both possess high-precision clocks, such
as Cs atomic clocks, running at exactly the same rate. They do not
agree on a common time at the same readout, for example twelve
o'clock. The difference of time $Y$ between their clocks can be
expressed as
\begin{align}
Y=t_B-t_A|_{\text{ Alice and Bob have the same readouts}}.
\end{align}
In a quantum scheme, in order to eliminate the relative phase that may emerge during
qubits transporting to the spatially separated locations, the
entangled states should be distributed to Alice and Bob
adiabatically. After the entanglement distribution, Alice and Bob
respectively perform measurements on all of their qubits
simultaneously when their clocks point to the same readout. We
choose a normalized entangled state
$|\psi\rangle=\sum_{i,j} p_{ij}|i_{A}\rangle |j_{B}\rangle$
which is merely changed with an overall unobservable phase under
the unitary evolution
$U_{AB}(t)=e^{-i\hat{H}_A t/\hbar}\otimes e^{-i\hat{H}_B t/\hbar}$.
$|i_{A}\rangle$ and $|j_{B}\rangle$ are orthonormal basis of
measurements which satisfy the completeness
$\sum_{i,j} |i_{A}\rangle|j_{B}\rangle\langle j_{B}|\langle i_{A}|=\mathbb{I}$
where $\mathbb{I}$ denotes to the identity. Suppose Alice performs
the measurement before Bob and obtains a result $|i_{A}\rangle$
with probability $P(i_{A})=\sum_j |p_{ij}|^2$, then the collapsed
state evolves as
\begin{eqnarray}
{e^{-\textrm{i}\hat{H}_{B}Y/\hbar}}\sum_j p_{ij}|j_{B}\rangle=\sum_{k,j}p_{ij}U_{kj}|k_{B}\rangle
\end{eqnarray}
where $U_{kj}=\langle k_{B}|e^{-i\hat{H}_{B}Y/\hbar}|j_{B}\rangle$.
Then Bob will obtain the result $|k_{B}\rangle$ with probability
$P(k_{B}|i_{A})= \frac{|\sum_j p_{ij}U_{kj}|^2}{P(i_{A})}$.
Thus, Fisher information theory and Crem\'{e}r-Rao bound \cite{fisher,cramer,holevo,Braunstein,caves,Escher} can be
utilized in this clock synchronization situation while the estimation
is asymptotically unbiased. By comparing the ratio of observed
measurement outcomes with probability distribution that is
determined by the parameter $Y$, two issues may prevent one estimating
$Y$ with a high precision. First, the number of experimental trials
is finite, so the ratio of measured outcomes may deviate from
the distribution; only when the number $\nu$ is large enough,
could the estimation be unbiased and the Cream\'{e}r-Rao bound be
reached. Second, a one-to-one mapping $P(\xi|Y)\leftrightarrow Y$
between the probability distributions and parameter is essential.
In this Letter,  we assume that Alice and Bob respectively
perform measurements
$\hat{X}=|\widetilde{0}\rangle\langle\widetilde{0}|-|\widetilde{1}\rangle\langle\widetilde{1}|$ on all of their own qubits simultaneously when their own clock
points to a specific value, where
$|\widetilde{x}\rangle=\frac{1}{\sqrt{2}}\sum_{i=0}^{1}(-1)^{x\cdot i}|i\rangle$
$(x=0,1)$, and
$|0\rangle,|1\rangle$ are the orthogonal eigenstates of each
qubit, which have the identical Hamiltonian $\hat{H}$ satisfying
$\hat{H}|0\rangle= E_0|0\rangle$, $\hat{H}|1\rangle=E_1|1\rangle$,
and $\omega=(E_1-E_0)/\hbar>0$. One way to implement these ticking
qubits in experiment is to put some spin-1/2 particles into the
magnetic fields with the same field strength.

\emph{Quantum clock synchronization with Bell state and GHZ state.--}
At first, we suppose that Alice and Bob share $N\nu$ pairs entangled
qubits with form $|\Psi^{(-)}\rangle=(|01\rangle-|10\rangle)/\sqrt{2}$ \cite{jozsa}
which is invariant under unitary evolution
$e^{-i\hat{H}_{A}t/\hbar}\otimes{e}^{-i\hat{H}_{B}t/\hbar}$.
Alice and Bob perform measurements expressed as an operator
$\hat{f}=\hat{X}(t_A)\otimes \hat{X}(t_B)$ in the Heisenberg picture,
and it can be described as a set of positive operator
valued measurements with element
$\hat{E}(\xi)=|\widetilde{x}\rangle\langle \widetilde{x}|\otimes|\widetilde{y}\rangle\langle\widetilde{y}|$
while $\xi=(\widetilde{x},\widetilde{y})$ where $x,y=0,1$.
Then, the probability distribution of all the measurement results is
\begin{equation}
P(\xi|Y)=\textrm{Tr}[\hat{E}(\xi)\hat{\rho}(Y)]=\frac{1}{2}\left(\delta_{x,-y} \cos^2\frac{\beta}{2} +\delta_{x,y}\sin^2 \frac{\beta}{2}\right)
\end{equation}
and the Fisher information is calculated as
\begin{equation}
\mathcal{F}_{Y}=\sum_{\xi}P(\xi|Y)\left[\frac{\partial\ln{P}(\xi|Y)}{\partial{Y}}\right]^2
=\omega^{2}
\end{equation}
where $\hat{\rho}(Y)$ is the density matrix of pure state
$e^{-i\hat{H}_{A}t/\hbar}\otimes e^{-i\hat{H}_{B}(t+Y)/\hbar}|\Psi^{(-)}\rangle$,
$\beta=\omega |Y|$, and $\delta_{x,y}$ is Kronecker's delta.
In addition the average of the measurement operator can be calculated
as $\langle \hat{f} \rangle_Y=\sum_{\xi} g(\xi) P(\xi|Y)=-\cos \beta$
where $g(\xi)=(-1)^{x+y}$.
If $|Y|<\pi/\omega$ holds, we can obtain the difference of time
$|Y_{est}|$ from the observed expectation value
$\overline{f}=\frac{1}{N\nu}\sum_{i=1}^{N\nu}g(\xi_i)$ after
measurements and classical communication. The sign of $Y_{est}$ can
be determined by the outcomes of Alice's and Bob's measurement because
the one who firstly performed $N\nu$ times measurement would get a duel
results with probabilities
$P(|\widetilde{0}\rangle)=P(|\widetilde{1}\rangle)=1/2$. Furthermore,
the uncertainty of estimation of $Y$ could reach the Cram\'{e}r-Rao bound
$\delta Y_{est}={1}/(\omega \sqrt{N\nu\mathcal{F}_{Y}})={1}/({\omega\sqrt{N\nu}})$
which is the standard quantum limit. Therefore, in this scheme Alice and Bob
can synchronize their clocks with accuracy ${1}/({\omega\sqrt{N\nu}})$.

Some researches have also focused on using the quantum entanglement
strategies and employing GHZ like states,
$|\textrm{GHZ}\rangle=(|0\rangle^{\otimes N}_A|1\rangle^{\otimes N}_B+|1\rangle^{\otimes N}_A|0\rangle^{\otimes N}_B)/{\sqrt{2}}$,
to enhance the precision of parameter estimation
\cite{bollinger,Giovannetti3,Giovannetti1,Giovannetti2}.
It is easy to verify that GHZ state is merely changed with an overall
phase under the unitary evolution
$(e^{-i\hat{H} t/\hbar})^{\otimes N}_A \otimes (e^{-i\hat{H} t/\hbar})^{\otimes N}_B $.
The probability distribution in this protocol takes the form:
$P(\xi'|Y)=\frac{1}{2^{2N}}[1+g'(\xi')\cos(N \beta)]$,
where $g'(\xi')=(-1)^{\sum_k (x_k+y_k)}$ while the symbol is
$\xi'=(\widetilde{x}_{1},\cdots,\widetilde{x}_{N},\widetilde{y}_{1},\cdots, \widetilde{y}_N)$,
$x_i,y_j=0,1$; and
$\hat{f}'=\hat{X}(t_A)^{\otimes N}\otimes \hat{X}(t_B)^{\otimes N}$
is the measurement operator in the Heisenberg picture. The average of the
operator is calculated as $\cos(N \beta)$,
and Fisher information $\mathcal{F}_{Y}$ is $N^2\omega^{2}$.
Considering that the probability
distributions and expectation value are all function with
periodicity $2\pi/N$, thus one can unambiguously obtain
$|Y_{est}|$ from the observed expectation value after measurements
and classical communication only when the condition
$|Y|<\pi/(\omega N)$ is satisfied. The sign of $Y_{est}$ can
also be determined by the outcomes of Alice's and Bob's measurement
because the one who firstly performed $N\nu$ times measurement would
get the probability
$P(|\widetilde{x_1}\cdots\widetilde{x_N}\rangle)={1}/{2^N}$
and $P(|\widetilde{x_i}\rangle)=1/2$.
When $\nu$ is large enough, the uncertainty can attain the
Cram\'{e}r-Rao bound
$\delta Y_{est}={1}/{\sqrt{\nu\mathcal{F}_{Y}}}={1}/({\omega N\sqrt{\nu}})$
which has a Heisenberg scaling accuracy. Despite of this optimal
local distinguishability in the Hilbert space, GHZ states are
inappropriate to obtain more advantageous information from any
values of parameter $Y$ in this single procedure since the condition
$|Y|<\pi/(\omega N)$ is required \cite{durkin}.

\begin{figure}[t]
\includegraphics[width=0.45\textwidth]{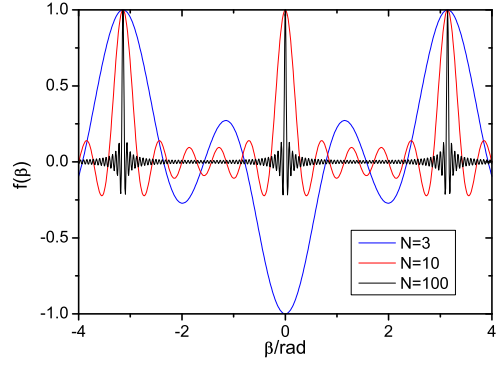}\\
\caption{The functional relation between the expectation value
$f(\beta)$ and the parameter $\beta$. }
\label{expectation}
\end{figure}

\emph{Quantum clock synchronization with bipartite maximally entangled states.---}
We next consider a scheme which exploit different entanglement resource.
The bipartite maximally entangled spin-zero singlet has been
proposed as a resource for quantum-enhanced metrology \cite{cable},
with the following form:
\begin{eqnarray}
|\chi\rangle=\frac{1}{\sqrt{2J+1}}\sum_{M=-J}^{J}(-1)^{J-M}|J,M\rangle_{z,A}|J,-M\rangle_{z,B}
\end{eqnarray}
where $J=N/2$, and $|J,M\rangle_z$ is a completely symmetric
normalized state (Dicke state) with $(J-M)$ qubits being $|0\rangle$
and $(J+M)$ qubits being $|1\rangle$. There is an explicit mapping
between the two-symmetric entangled state and the direct product
of $N$ maximally entangled states, which is presented in
\cite{PhysRevA.84.034302}, then one obtains
\begin{eqnarray}
|\chi\rangle&=&\frac{2^{N/2}}{\sqrt{N+1}}{\mathbb{I}}^{\otimes N}\otimes S|\Psi^{(-)}\rangle^{\otimes N}\nonumber\\
&=&\frac{2^{N/2}}{N!\sqrt{N+1}}\sum_{\sigma}|\Psi^{(-)}\rangle_{A_1B_{\sigma_1}} \cdots |\Psi^{(-)}\rangle_{A_N B_{\sigma_N}}
\end{eqnarray}
where ${\mathbb{I}}$ is the identity operator on Hilbert space
$\mathcal{H}=\{|0\rangle,|1\rangle\}$,
$S=\sum_{M=-J}^J |J,M\rangle_z \langle J,M|$ is the symmetric
projector that maps states in $\mathcal{H}^{\otimes N}$ onto its
symmetric subspace $\mathcal{H}^{\otimes N}_+$, $\sigma$ denotes to
a permutation. After the adiabatic distribution of $N\nu$ pairs
entanglement $|\Psi^{(-)}\rangle$, Alice or Bob can perform the
symmetric projector $S$ to obtain $\nu$ pairs $|\chi\rangle$.
Because $|\Psi^{(-)}\rangle=(|01\rangle-|10\rangle)/\sqrt{2}$
changes only with an overall unobservable phase under any unitary
evolution of form $U\otimes{U}$ in two-qubit space, then this
singlet has the rotational invariance property under unitary evolution
$U^{\otimes N}\otimes U^{\otimes N}$ and has  identical
expression in any spin basis, e.g. $z\mapsto x \mapsto y$ when
ignoring the overall phase. Another proof has been presented in
Ref.~\cite{Phys. Rev. A 72.012307}, and this invariance property has
been tested in the experiment \cite{six-photon}. These bipartite
maximally entangled states play an important role in the quantum
information distribution and concentration \cite{PhysRevA.59.156,PhysRevA.61.032311,zhang}. Recently these
states used in our scheme have been generated experimentally by
using stimulated parametric down-conversion and have been used
in the $1$ to $3+2$ information distribution
\cite{six-photon,telecloning}. Additionally, some other experiments
also produce such entanglement and realize the quantum information
distribution \cite{Phys.Rev.A 82.030302}. Next, we show
that these technologies can also be utilized to implement our scheme
for quantum clock synchronization. The pure state evolved as
\begin{eqnarray}
&&(e^{-i\hat{H}_{A}t_{A}/\hbar})^{\otimes N}\otimes(e^{-i\hat{H}_{B}t_{B}/\hbar})^{\otimes N}|\chi\rangle\nonumber\\
&=&(\mathbb{H}_{A}^{\otimes N}\otimes\mathbb{H}_{B}^{\otimes N})(\mathbb{I}_{A}^{\otimes N}\otimes(e^{-iY\mathbb{H}_{B}\hat{H}_{B}\mathbb{H}_{B}/\hbar})^{\otimes N})|\chi\rangle\nonumber\\
&=&(\mathbb{H}_{A}^{\otimes N}\otimes\mathbb{H}_{B}^{\otimes N})(\mathbb{I}_{A}^{\otimes N}\otimes{U}_{B}(\pi/2,\beta,-\pi/2)^{\otimes N})|\chi\rangle
\end{eqnarray}
where the overall phase is ignored;
$\mathbb{H}_{A,B}=\frac{1}{\sqrt{2}}\begin{bmatrix}1&1\\1&-1\end{bmatrix}$
is the Hadmard matrix and the unitary operator $U(\alpha,\beta,\gamma)$
is expressed by three Euler angles in the basis
$\{|0\rangle, |1\rangle \}$ as the following:
\begin{eqnarray}
U(\alpha,\beta,\gamma)=\left[
\begin{array}{c c}
\cos\frac{\beta}{2} e^{i(\alpha+\gamma)/2} & \sin\frac{\beta}{2} e^{-i(\alpha-\gamma)/2} \\
-\sin\frac{\beta}{2} e^{i(\alpha-\gamma)/2}& \cos\frac{\beta}{2} e^{-i(\alpha+\gamma)/2}
\end{array}\right].
\end{eqnarray}
Moreover, according to group theory of the irreducible
representation, we obtain an analytical expression of the
unitary operator in $N$-qubit space \cite{group}
\begin{eqnarray}
U(\alpha,\beta,\gamma)^{\otimes N}|JM\rangle=\sum_{M'}e^{-i(M\alpha+M'\gamma)}d^{J}_{M',M}(\beta)|JM'\rangle.
\end{eqnarray}
Thus, we can obtain the probability distribution of measurement
outcomes $\xi'=(\widetilde{x}_{1},\cdots,\widetilde{x}_{N},\widetilde{y}_{1},\cdots, \widetilde{y}_N)$, with $x_i,y_j=0,1$;
\begin{eqnarray}
P(\xi'|Y)=\frac{\left[d^J_{M',-M}(\beta)\right]^2}{(2J+1)C^{J-M}_{2J}C^{J-M'}_{2J}}.
\end{eqnarray}
where $(J-M)$ is the number of $0$ in $\{x_1,\cdots, x_N\}$
while $(J-M')$ is the number of $0$ in $\{y_1,\cdots, y_N\}$.
As before the expectation value of measurement operator
$\hat{f}'=\hat{X}(t_A)^{\otimes N}\otimes \hat{X}(t_B)^{\otimes N}$
is calculated as
\begin{eqnarray}
f(\beta):=\langle\hat{f'}\rangle_{Y}&=&\sum^{J}_{M,M'=-J}(-1)^{N+M+M'}\frac{[d^J_{M',-M}(\beta)]^2}{2J+1}\nonumber\\
&=&\frac{(-1)^N}{N+1}\frac{\sin(N+1)\beta}{\sin \beta}.
\end{eqnarray}
The function $f(\beta)$ against its argument $\beta$ is shown in
Fig.~\ref{expectation} with different numbers of qubits.
Furthermore, Fisher information reads
\begin{eqnarray}
\mathcal{F}_{Y}=\sum_{M,M'=-J}^J\frac{4\left[\frac{d }{dY}d^J_{M',-M}(\beta)\right]^2}{2J+1}=\frac{4J(J+1)\omega^{2}}{3}
\end{eqnarray}
with which it is straightforward to find that lower bound
$\delta Y_{est}=\sqrt{3}/(\omega\sqrt{N(N+2)\nu})$
obviously breaks the quantum standard limit and performs a Heisenberg
scaling accuracy.

Nevertheless, this scheme has some special properties that differ from
the previous schemes. Considering that $f(\beta)$ is clearly ``peaked"
around $\beta=0$ with width $\sim{\pi}/{(N+1)}$ (see
Fig.~\ref{expectation}), we can confirm that the uncertainty of
$Y_{est}$ could reach the Cram\'{e}r-Rao bound
${1}/({\omega\sqrt{\nu\mathcal{F}_{Y}}})$ for a large number $\nu$
when the condition $|Y|<\pi/((N+1)\omega)$ is satisfied.
Although the
bipartite maximally entangled spin-zero singlet fails to gain more
advantageous information of the parameter $Y$ from the observed
expectation value against the GHZ state, more
importantly we will acquire a quantum enhanced criterion to judge
whether two remote clocks have been synchronized with the accuracy
$\pi/(\omega(N+1))$ even when the number $\nu$ is not large enough,
i.e., the unbiased estimation hypothesis is not fulfilled. When
the expectation value
$\overline{f}=\frac{1}{\nu}\sum_{i=1}^{\nu} g(\xi_i)$
obtained from the outcomes of experiments after classical
communication satisfies
$|\overline{f}-(-1)^N|\leq 1-{1}/{\sqrt{2}}$, one obtains
\begin{eqnarray}
&&Prob\left(|Y|\leq\frac{\pi}{(N+1)\omega}\right)\gtrsim Prob\left(\left|\overline{f}-\langle\hat{f}\rangle_{Y}\right|\leq\frac{1}{\sqrt{2}}\right)\nonumber\\
&&\geq1-2e^{-\frac{\nu}{4}}
\end{eqnarray}
which is Chernoff bound \cite{nielsen}. For example, suppose that
$|\overline{f}-(-1)^N|\leq 1-{1}/{\sqrt{2}}$ and $\nu=10$, we can
infer that the inequality $|Y|\leq{\pi}/({(N+1)\omega})$ holds with
fiducial probability larger than $84\%$.

\emph{Conclusions.---}
In many cases of quantum metrology, quantum Cram\'{e}r-Rao
bound can only be achieved asymptotically ,i.e., it holds
for unbiased estimation with  infinite number $\nu$ and zero
error $\delta Y_{est}\rightarrow0$ \cite{zzb}; and unfortunately
this problem also exists in some quantum clock synchronization
strategies. However, applying the bipartite maximally
entangled spin-zero singlet one can obtain a standard to judge
whether two spatially separated clocks have been synchronized
within a specific uncertainty even when the number $\nu$ is not
large enough. If so, we can step further to obtain the difference
between two clocks with a Heisenberg scaling accuracy in accordance
to the expectation of the measurement results by increasing the
number $\nu$.

In conclusion, we propose a novel quantum scheme for remote clocks synchronization within
a specific accuracy. This bound of accuracy scales as the Heisenberg limit
which is the ultimate limit of precision measurements under all conditions.
With developments in creating experimentally the entanglement resources,
this quantum scheme of remote clocks synchronization may be implemented
which may possess unprecedent precision.

We would like to thank Guo-Yong Xiang, Xiang-Ru Xiao and Li Jing
for useful discussions. This work was supported by 973 program (2010CB922904),
NSFC (11175248), NFFTBS(J1030310,J1103205), grants from Chinese Academy of Sciences,
and the Chun-Tsung scholar fund of Peking University.


\begin{thebibliography}{99}
\bibitem{einstein} A. Einstein, Ann. Phys. \textbf{17}, 891 (1905).
\bibitem{eddington} A. S. Eddington, \emph{The Mathematical Theory of Relativity}, 2nd edition, (Cambridge University Press, Cambridge, 1924).
\bibitem{jozsa} R. Jozsa \emph{et al.}, Phys. Rev. Lett. \textbf{85}, 2010 (2000);
E. A. Burt \emph{et al.}, Phys. Rev. Lett. \textbf{87}, 129801(2001);
R. Jozsa \emph{et al.}, Phys. Rev. Lett. \textbf{87}, 129802 (2001).
\bibitem{krco} M. Krco, and P. Paul, Phys. Rev. A \textbf{66}, 024305 (2002).
\bibitem{ben-av} R. Ben-Av, and I. Exman, Phys. Rev. A \textbf{84}, 014301 (2011).
\bibitem{ren} C. Ren, and H. F. Hofmann, Phys, Rev. A \textbf{86}, 014301 (2012).
\bibitem{APL} A. Valencia \emph{et al.}, Appl. Phys. Lett. \textbf{85}, 2655 (2004).
\bibitem{mpe} Z. Zhao \emph{et al.}, Science \textbf{430}, 54 (2004);
X. C. Yao \emph{et al.}, Nat. Photon. \textbf{6}, 225 (2012).
\bibitem{Phys. Rev. D 23.1693} C. M. Caves, Phys. Rev. D \textbf{23}, 1693 (1981).
\bibitem{fisher} R. A. Fisher, Proc. Camb. Soc. \textbf{22}, 700 (1925).
\bibitem{cramer} H. Cram\'{e}r, \emph{Mathematical Methods of Statistics} (Princeton University, Princeton, NJ, 1946).
\bibitem{holevo} A. S. Holevo, \emph{Probabitistic and Statistica Aspects of Quantum Theory} (North-Holland, Amsterdam, 1982), especially Chaps. III.2 and VI.2.
\bibitem{Braunstein} S. L. Braunstein, J. Phys. A \textbf{25}, 3813 (1992); S. L. Braunstein, Phys. Rev. Lett. \textbf{69}, 3598 (1992).
\bibitem{caves} S. L. Braunstein, and C. M. Caves, Phys. Rev. Lett. \textbf{72}, 3439 (1994).
\bibitem{Escher} B. M. Escher \emph{et al.}, Nat. Phys. \textbf{7}, 406 (2011).
\bibitem{bollinger} J. J. Bollinger, W. M. Itano, D. J. Wineland and D. J. Heinzen
 Phys. Rev. A \textbf{54}, R4649 (1996).
\bibitem{Giovannetti3} V. Giovannetti, S. Lloyd, and L. Maccone , Science \textbf{306}, 1330 (2004).
\bibitem{Giovannetti1} V. Giovannetti, S. Lloyd, and L. Maccone, Phys. Rev. Lett. \textbf{96}, 010401 (2006).
\bibitem{Giovannetti2} V. Giovannetti, S. Lloyd, and L. Maccone, Nat. Photon. \textbf{5}, 222 (2011).
\bibitem{Nagata} T. Nagata, R. Okamoto, J. L. O¡¯Brien, K. Sasaki, and S. Takeuchi, Science \textbf{316}, 726 (2007).
\bibitem{higgins} B. L. Higgins \emph{et al.}, Nature. \textbf{450}, 393 (2007).
\bibitem{Kacprowicz} M. Kacprowicz \emph{et al.}, Nat. Photon. \textbf{4}, 357 (2010).
\bibitem{afek} I. Afek, O. Ambar, and Y. Silberberg, Science \textbf{328}, 879 (2010).
\bibitem{xiang} G. Y. Xiang \emph{et al.}, Nat. Photon. \textbf{5}, 43 (2011).
\bibitem{Yonezawa} H. Yonezawa \emph{et al.}, Science \textbf{337}, 1514 (2012).
\bibitem{chuang} I. L. Chuang, Phys. Rev. Lett. \textbf{85}, 2006 (2000).
\bibitem{burgh} M. de Burgh and S. D. Bartlett, Phys. Rev. A \textbf{72}, 042301 (2005).
\bibitem{durkin} G. A. Durkin, and J. P. Dowling, Phys. Rev. Lett. \textbf{99}, 070801 (2007).
\bibitem{cable} H. Cable, and G. A. Durkin, Phys. Rev. Lett. \textbf{105}, 013603 (2010).
\bibitem{PhysRevA.84.034302} Y. N. Wang, H. D. Shi, Z. X. Xiong, L. Jing, X. J. Ren, L. Z. Mu, and H. Fan, Phys. Rev. A \textbf{84}, 034302 (2011).
\bibitem{Phys. Rev. A 72.012307} J. Schliemann, Phys. Rev. A \textbf{72}, 012307 (2005).
\bibitem{six-photon} M. R{\aa}dmark, Marek \.{Z}ukowski, and M. Bourennane, Phys. Rev. Lett. \textbf{103}, 150501 (2009).
\bibitem{PhysRevA.59.156} M. Murao, D. Jonathan, M. B. Plenio, and V. Vedral, Phys. Rev. A \textbf{59}, 156 (1999).
\bibitem{PhysRevA.61.032311} M. Murao, M. B. Plenio, and V. Vedral, Phys. Rev. A \textbf{61}, 032311 (2000).
\bibitem{zhang} Y. L. Zhang \emph{et al.},  Phys. Rev. A \textbf{87}, 022302 (2013).
\bibitem{telecloning}M. R{\aa}dmark, M. Zukowski and M. Bourennane, New J. Phys. \textbf{11}, 103016 (2009); M. R{\aa}dmark, M. Wiesniak, M. Zukowski, and M. Bourennane , Phys. Rev. A \textbf{80}, 040302(R) (2009).
\bibitem{Phys.Rev.A 82.030302} A. Lamas-Linares, J. C. Howell, and D. Bouwmeester,
Nature (London) \textbf{412}, 887 (2001); F. Ciccarello, M.
Paternostro, S. Bose, D. E. Browne, G. M. Palma, and
M. Zarcone, Phys. Rev. A \textbf{82}, 030302(R) (2010); H.
Yu, Y. Luo, and W. Yao, Phys. Rev. A \textbf{84}, 032337 (2011);
A. Chiuri, C. Greganti, M. Paternostro, G. Vallone, and
P. Mataloni, Phys. Rev. Lett. \textbf{109}, 173604 (2012).
\bibitem{group} Z. Ma, \emph{Group Theory for Physicists} (World Scientific, Singapore, 2007);\\
$d^J_{M',M}(\beta)=\sum_\nu(-1)^\nu\frac{[(J+M')!(J-M')!(J+M)!(J-M)!]^{1/2}}{(J+M'-\nu)!(J-M-\nu)!\nu!(\nu+M-M')!}
\times\\
(\cos\frac{\beta}{2})^{2J+M'-M-2\nu}(\sin\frac{\beta}{2})^{2\nu-M'+M}$.
\bibitem{nielsen} A. S. Nelsen, and I. L. Chuang, \emph{Quantum Computation and Quantum Information} (Cambridge University Press, Cambridge, England, 2000).
\bibitem{zzb} M. Tsang, Phys. Rev. Lett. \textbf{108}, 230401 (2012);
D. W. Berry \emph{et al.}, Phys. Rev. A \textbf{86}, 053813 (2012).
\end{thebibliography}
\end{document}